\documentclass{ieeetmlcn}
\usepackage{multirow}
\usepackage{cite}
\usepackage{amsmath,amssymb,amsfonts}
\usepackage{algorithmic}
\usepackage{graphicx,color}
\usepackage{textcomp}
\usepackage{xcolor}
\usepackage{hyperref}
\hypersetup{hidelinks=true}
\usepackage{algorithm,algorithmic}
\def\BibTeX{{\rm B\kern-.05em{\sc i\kern-.025em b}\kern-.08em
    T\kern-.1667em\lower.7ex\hbox{E}\kern-.125emX}}
\AtBeginDocument{\definecolor{tmlcncolor}{cmyk}{0.93,0.59,0.15,0.02}\definecolor{NavyBlue}{RGB}{0,86,125}}

\def\OJlogo{\vspace{-4pt}\includegraphics[height=18pt]{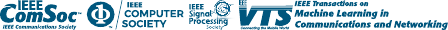}}
\def\seclogo{\vspace{10pt}\includegraphics[height=18pt]{new_logo}}

\def\authorrefmark#1{\ensuremath{^{\textbf{#1}}}}

\begin{document}

\markboth{}{Murshed {et al.}}

\title{MetaFAP: Meta-Learning for Frequency Agnostic Prediction of Metasurface Properties}

\author{Rafid Umayer Murshed\authorrefmark{1}, Md Shoaib Akhter Rafi\authorrefmark{2}, Sakib Reza\authorrefmark{1}, Mohammad Saquib\authorrefmark{1}, and Ifana Mahbub\authorrefmark{1}}

\affil{Department of Electrical and Computer Engineering, The University of Texas at Dallas, Richardson, TX, USA}
\affil{Department of Electrical and Electronic Engineering, Bangladesh University of Engineering and Technology, Dhaka, Bangladesh}

\corresp{Corresponding author: Rafid Umayer Murshed (email: rafidumayer.murshed@utdallas.edu).}

\authornote{}

\begin{abstract}
Metasurfaces, and in particular reconfigurable intelligent surfaces (RIS), are revolutionizing wireless communications by dynamically controlling electromagnetic waves. Recent wireless communication advancements necessitate broadband and multi-band RIS, capable of supporting dynamic spectrum access and carrier aggregation from sub-6\,GHz to mmWave and THz bands.  The inherent frequency dependence of meta-atom resonances degrades performance as operating conditions change, making real-time, frequency-agnostic metasurface property prediction crucial for practical deployment. Yet, accurately predicting metasurface behavior across different frequencies remains challenging. Traditional simulations struggle with complexity, while standard deep learning models often overfit or generalize poorly. To address this, we introduce MetaFAP (\emph{Meta}-Learning for \emph{F}requency-\emph{A}gnostic \emph{P}rediction), a novel framework built on the meta-learning paradigm for predicting metasurface properties. By training on diverse frequency tasks, MetaFAP learns broadly applicable patterns. This allows it to adapt quickly to new spectral conditions with minimal data, solving key limitations of existing methods. Experimental evaluations demonstrate that MetaFAP reduces prediction errors by an order of magnitude in MSE and MAE while maintaining high Pearson correlations. Remarkably, it achieves inference in less than a millisecond, bypassing the computational bottlenecks of traditional simulations, which take minutes per unit cell and scale poorly with array size. These improvements enable real-time RIS optimization in dynamic environments and support scalable, frequency-agnostic designs. MetaFAP thus bridges the gap between intelligent electromagnetic systems and practical deployment, offering a critical tool for next-generation wireless networks.
\end{abstract}



\maketitle
\section{Introduction}
\IEEEPARstart{M}{etasurfaces} have emerged as a transformative technology in wireless communications, offering unprecedented control over the radio environment through the dynamic manipulation of electromagnetic waves \cite{DiRenzo2020_JSAC}. In particular, reconfigurable intelligent surfaces (RIS) exploit sub-wavelength engineered meta-atoms to tailor the phase, amplitude, and polarization of incident signals \cite{Liaskos2018_CommMag, Tapio2021_EURASIP}. This capability enables critical applications—including passive beamforming, interference suppression, energy-efficient propagation, and enhanced physical-layer security—that are essential for meeting the stringent performance requirements of both 5G Advanced and emerging 6G networks \cite{ Dai2020_Access, Faisal2022_Access}.

Despite the significant benefits, designing an effective RIS remains challenging due to the complex electromagnetic (EM) behavior of individual meta-atoms. The EM response—characterized by reflectance, transmittance, and absorptivity—is highly sensitive to variations in geometry, material properties, incident angles, and operating frequencies \cite{Costa2021_OJComSoc}. Traditional design methodologies rely heavily on full-wave EM simulations and analytical circuit models \cite{Berto2024_EuMC}. While full-wave solvers offer high accuracy by rigorously solving Maxwell’s equations, they are computationally expensive and impractical for exhaustive exploration or real-time optimization \cite{Tapio2021_EURASIP}. On the other hand, Analytical models offer faster evaluations by approximating metasurfaces as transmission line circuits or impedance sheets. However, they rely on simplifying assumptions that limit their accuracy over wide operating ranges, especially when incident angles and frequencies vary \cite{Costa2021_OJComSoc, Berto2024_EuMC}.

To mitigate these challenges, recent research has increasingly turned to data-driven approaches using machine learning (ML) and deep learning (DL) to develop surrogate models that can rapidly predict RIS behavior \cite{Liu2023_arXiv}. Supervised neural networks have been successfully employed to approximate the nonlinear mapping from metasurface configurations (including geometry, materials, and bias states) to their EM responses \cite{Ghorbani2021_SciRep}. These networks deliver predictions in milliseconds and significantly accelerate the design cycle. Nevertheless, these conventional ML/DL methods are highly data-dependent and tend to perform poorly when extended beyond the frequency bands or design conditions present in their training sets. For example, a model trained on a narrow frequency band may fail to accurately predict metasurface behavior when applied to an unseen frequency range, a critical shortcoming in dynamic wireless environments.

The demand for broadband or multi-band functionality further exacerbates this issue. While many metasurfaces are traditionally engineered for a single frequency band, next-generation applications increasingly require RIS that operate efficiently across diverse spectral regimes. These regimes range from sub-6 GHz to mmWave and even THz bands \cite{lan2023real}. This expanded frequency coverage is crucial for advanced wireless systems that utilize dynamic spectrum access, carrier aggregation, and spectrum sharing to meet ever-growing capacity and latency requirements \cite{Wu2020_CommMag,DiRenzo2020_JSAC}. However, each meta-atom’s resonance is frequency dependent. A design optimal at 5 GHz may underperform at 28 GHz. This frequency shift may cause unpredictable phase errors and reduce beamforming gains.\cite{Ghorbani2021_SciRep}. 

Moreover, real-time reconfiguration is becoming indispensable as RIS-based networks must adapt to changing user demands, environmental conditions, and regulatory constraints \cite{reconfigure_realtime,wen2023real}. In such scenarios, redesigning or recalibrating a metasurface is impractical. Full-wave simulations are computationally expensive. This is especially true for large-scale deployments with hundreds or thousands of meta-atoms \cite{mansouree2021large}. Instead, predictive models must be capable of \emph{extrapolating} beyond the conditions observed during training or calibration. Achieving this level of frequency agility calls for fast, accurate, and robust modeling techniques. These techniques must generalize to unseen frequency bands, which ensures RIS operates effectively across wide bandwidths without frequent retraining. Recent deep-learning design methods show promise over limited spectral ranges. However, their accuracy drops outside the trained distribution. This highlights the need for more flexible, \emph{frequency-agnostic} approaches \cite{Ghorbani2021_SciRep,Ji2023_Light}.

Motivated by these challenges, our work explores the potential of meta-learning—“learning to learn”—to overcome the generalization limitations of conventional models. Meta-learning frameworks, such as Model-Agnostic Meta-Learning (MAML) \cite{Finn2017_ICML, Nichol2018_arXiv}, aim to learn an initialization that can be rapidly adapted to new tasks with only a few training examples. In our context, each frequency band or RIS configuration is treated as a distinct task. It enables the meta-learned model to be fine-tuned quickly to accommodate new spectral conditions with minimal additional data. This approach endows our predictive model with “prior knowledge” that enhances its ability to generalize beyond the training distribution—a crucial capability for real-time network adaptation where rapid reconfiguration is imperative. In dynamic scenarios involving user mobility or rapid environmental changes, an RIS that can adjust its electromagnetic response nearly in real time can significantly enhance link reliability and throughput \cite{Wu2020_CommMag}. Moreover, by reducing the need for extensive field measurements and iterative tuning, our method offers a scalable, frequency-agnostic solution for next-generation wireless environments.

In summary, our work bridges the gap between high-fidelity EM modeling and practical RIS design by integrating meta-learning into the predictive framework. To the best of our knowledge, this is the first study to deploy a meta-learning-based approach for metasurface property prediction. By leveraging MAML to model the reflectance and transmittance characteristics of RIS elements across diverse frequency ranges, our approach accelerates the design cycle and reduces computational overhead. It also robustly generalizes to unseen spectral conditions. This paves the way for adaptive and scalable RIS deployments in dynamic wireless networks \cite{DiRenzo2020_JSAC, Wu2020_CommMag, Finn2017_ICML}.

The primary contributions of this paper are summarized as follows:
\begin{enumerate}

\item We introduce MetaFAP, the first meta-learning framework leveraging Model-Agnostic Meta-Learning (MAML), specifically designed to predict metasurface properties (reflectance, transmittance,  and absorbance) in a frequency-agnostic manner. This approach significantly addresses the limitations of existing deep-learning models, which struggle to generalize beyond trained frequency bands.
 \item Our experimental results demonstrate that MetaFAP achieves an order-of-magnitude improvement in prediction accuracy compared to traditional ML and DL methods. Notably, it attains a mean squared error (MSE) as low as 0.0079 and maintains Pearson correlation coefficients around 80\%, substantially surpassing the state-of-the-art.

\item MetaFAP significantly enhances computational efficiency, reducing inference times to only 0.15 ms per prediction and requiring less than 30 minutes for training on an NVIDIA A100 GPU. Unlike full-wave electromagnetic simulations, which scale poorly with metasurface complexity, MetaFAP provides constant inference time regardless of array size.

\item Through an extensive ablation study, we identify and quantify the critical roles of frequency-dependent feature initialization and rapid adaptation mechanisms inherent in the meta-learning architecture. These insights underline why our framework outperforms conventional methods in adapting quickly and accurately to new frequency conditions.

\item We provide compelling practical scenarios where MetaFAP offers substantial benefits. These include adaptive beamforming in frequency-agile 6G urban networks, rapid real-time reconfiguration for vehicle-to-everything (V2X) and IoT applications. MetaFAP thus represents a significant step toward practical, scalable, and adaptive metasurface implementations in dynamic wireless communication environments.

\end{enumerate}

The remainder of the paper is organized as follows: Section II reviews the related work on conventional metasurface design, data-driven surrogate modeling, and meta-learning approaches across various domains. Section III presents the background on MAML and formulates the problem of predicting metasurface properties as a supervised learning task. Section IV describes our proposed methodology, including dataset generation, the architectural overview of our base model, and the detailed MAML framework. Section V provides an in-depth experimental evaluation, including comparisons with state-of-the-art models, analysis of performance under varying data conditions, and ablation studies. Finally, Section VI discusses the implications of our findings, and Section VII concludes the paper while outlining potential future research directions.

\section{Related Works}

The evolution of metasurface design has followed a clear trajectory. It began with traditional physics-based approaches, advancing to data-driven ML/DL methodologies, and most recently exploring other advanced learning techniques to overcome inherent limitations. This section reviews these developments, highlighting both their achievements and the challenges that motivate our work.

\subsection{Traditional Approaches} Conventional metasurface design and analysis relies heavily on physics-based models and EM simulations to predict and optimize key performance metrics such as reflection, refraction, absorption, and transmission \cite{Faisal2022_Access}. Early works employed rigorous full-wave EM simulations and analytical models to define these tasks. For example, Berto et al. \cite{Berto2024_EuMC} designed four sub-wavelength unit-cells at 30 GHz using three metal layers and two dielectric substrates to achieve 2-bit quantized transmission and reflection. This work demonstrated the practical feasibility of tailoring metasurface responses via full-wave simulation. Similarly, Suzuki et al. \cite{Suzuki2024_APS} developed a novel RIS element that incorporated a PIN diode to enable precise beamforming by minimizing vertical current components that degrade cross-polarization discrimination (XPD). In parallel, Costa and Borgese \cite{Costa2021_OJComSoc} introduced simplified yet accurate analytical circuit models. These models represented metasurfaces as transmission line circuits that capture critical parameters such as incident angle, mutual coupling, and ground plane interactions. Further contributions by Ndjiongue \cite{Ndjiongue2021_WCM} optimized metasurface parameters—specifically, determining the optimal unit-cell depth for minimizing interference between incoming and reflected waves using a liquid crystal on silicon (LCoS) based design.

Building on these foundational studies, full-wave EM simulations using tools like CST Microwave Studio, Ansys HFSS, or FDTD codes have become the gold standard for design verification. For instance, Berto et al. \cite{Berto2024_EuMC} performed extensive parameter sweeps to achieve a desired 2-bit phase quantization at 30 GHz. In contrast, Suzuki et al. \cite{Suzuki2024_APS} utilized full-wave simulations, along with measurements, to validate RIS performance improvements in 5G scenarios. Despite their accuracy, these simulation-based methods are computationally expensive and time-intensive. This is especially true when exploring large design spaces or conducting frequency sweeps. To alleviate this, researchers have developed analytical and equivalent-circuit models. These models treat metasurfaces as arrays of resonant circuits, using transmission line or impedance network analogies to approximate reflection and transmission characteristics \cite{DiRenzo2020_JSAC}. Costa and Borgese \cite{Costa2021_OJComSoc} demonstrated that such models can yield closed-form expressions for reflection amplitude and phase, significantly reducing reliance on iterative simulations. Additionally, alternative strategies, such as using generalized sheet transition conditions (GSTCs) to model metasurfaces as homogeneous impedance sheets, have been proposed—albeit typically under narrowband assumptions. Complementing these methods, optimization-based theoretical design approaches have emerged. Tretyakov \cite{Tretyakov2015_PhilTrans} derived generalized reflection/transmission laws using surface polarizabilities, while Jian et al. \cite{Jian2021_ICCC} explored standardized design approaches for RIS. Notably, Ndjiongue et al. \cite{Ndjiongue2021_WCM} optimized metasurface depth to maximize received power at a specific incidence angle, and Dajer et al. \cite{Dajer2022_DigCommNet} applied optimization strategies to model channels in RIS-enabled networks. Despite their successes, traditional approaches are limited by scalability, computational efficiency, and generalizability—challenges that become particularly acute in broadband and multi-band scenarios.

\subsection{ML/DL-based Works} To overcome the limitations of traditional methods, researchers have increasingly turned to ML and DL for metasurface design and property prediction. These data-driven approaches capture the complex, nonlinear relationships between metasurface configurations and their EM responses. It enables rapid predictions that replace computationally intensive simulations \cite{Jian2021_ICCC,Ghorbani2021_SciRep}. Supervised neural networks, for example, have been used to approximate scattering parameters—such as reflection amplitude and phase—based on geometric features and frequency, yielding surrogate models that deliver predictions in milliseconds \cite{Peurifoy2018_SciAdv}. These techniques have been successfully applied in diverse domains ranging from nanoparticle design in photonics to RF metasurface characterization\cite{Liu2018_ACSPhotonics}. Moreover, inverse design strategies employing generative models (including GANs and variational autoencoders) have enabled automated exploration of design spaces, often matching or exceeding expert-designed structures \cite{Ma2021_NatPhoton,Song2022_arXiv}. Complementary approaches, such as physics-informed neural networks (PINNs) integrated with topology optimization, have further reduced data requirements by embedding Maxwell’s equations into the learning process \cite{Ji2023_Light}.

Nevertheless, conventional ML/DL methods face significant challenges. They typically require extensive training datasets—often comprising tens of thousands of samples—to achieve high accuracy, and are effective primarily within the frequency bands represented in the training data \cite{Liu2018_ACSPhotonics}. This narrow-band focus hampers generalization when metasurfaces are deployed in dynamic or multi-band environments. In such scenarios, operating frequencies may shift (e.g., from 3.5 GHz to 28 GHz) or new configurations emerge, leading to out-of-distribution errors \cite{JZhang2021_ICC,Sejan2022_Sensors}. Moreover, the inherent “black-box” nature of deep networks limits the physical insight into design parameters and their corresponding EM responses. At the same time, the computational expense of retraining models for new conditions further restricts their practical application \cite{Liu2018_ACSPhotonics,Park2021_TSP}.

\subsection{Meta-Learning in Various Domains} The limitations of conventional ML/DL approaches motivate us to explore meta-learning, which aims to endow predictive models with rapid adaptability to new tasks or operating conditions. Meta-learning has increasingly influenced a wide array of fields, including wireless communications and materials science, thereby demonstrating its potential for applications analogous to metasurface prediction. In wireless communications, meta-learning techniques have been applied to tasks such as channel estimation, signal detection, and beamforming. For instance, Park et al. \cite{Park2019_SPAWC} implemented a meta-learning approach for signal demodulation in IoT scenarios by treating each device’s channel as a distinct task. In contrast, Zhang et al. \cite{JZhang2021_ICC} employed meta-learning for channel prediction in massive MIMO systems, enabling rapid adaptation with minimal feedback data. Adaptive beamforming solutions based on meta-learning have demonstrated robust performance under channel variations while reducing computational overhead \cite{JZhang2021_ICC,Ge2021_TVT}. Moreover, meta-reinforcement learning strategies have been utilized for adaptive resource allocation in O-RAN networks, further highlighting the versatility of meta-learning in nonstationary environments \cite{Zhao2024_OJCS}.

In materials science, meta-learning has proven valuable for predicting material properties under data-scarce conditions. Chen et al. \cite{Chen2024_arXiv} introduced a meta-learning framework for forecasting properties such as the creep rupture life of alloys. This is particularly useful when experimental data collection is expensive. Analogous applications in chemistry include using few-shot graph neural networks with meta-learning for molecular property prediction \cite{Liu2018_ACSPhotonics}. This mirrors the challenges in predicting diverse metasurface responses. In the photonics community, transferable models have been proposed to address gaps in model generalization for metamaterials, as reviewed by Cerniauskas et al. \cite{Cerniauskas2024_Oxford}.

Beyond these application domains, meta-learning has seen remarkable progress in fields such as image classification, scene classification, data mining, and natural language processing \cite{song2023comprehensive}. Architectures such as the Simple Neural Attentive Learner (SNAIL) leverage temporal convolutions and causal attention to effectively incorporate historical context \cite{Mishra2018_arXiv}. Moreover, Model-Agnostic Meta-Learning (MAML) \cite{Finn2017_ICML} and its variants—including Hessian-Free MAML \cite{fallah2020convergence}, Bayesian MAML \cite{Yoon2018_NeurIPS}, alpha-MAML \cite{DBLP:journals/corr/abs-1905-07435}, Reptile \cite{Nichol2018_arXiv}, and FOMAML \cite{Finn2017_ICML, Nichol2018_arXiv}—have demonstrated their effectiveness across diverse tasks. Recent studies further highlight meta-learning’s adaptability through applications in e-commerce demand prediction \cite{xu2024f}, image segmentation \cite{hendryx2019meta}, and long-document text summarization \cite{maurya2024nlp}. In wireless communications, Zou et al. \cite{zou2021meta} employed a MAML-based approach to optimize phase shifts and power allocation in IRS-assisted NOMA networks, showcasing rapid adaptation with minimal data requirements.

Overall, traditional ML/DL approaches have advanced metasurface design and prediction. However, they are limited in scalability and computational efficiency. They also struggle to generalize across dynamic conditions. These limitations have driven the adoption of meta-learning. By integrating meta-learning for fast adaptation to new frequency bands and design configurations, our work seeks to bridge these gaps.

\section{Background and Problem Formulation}

\subsection{Background on MAML}
The field of ML/DL has experienced a significant change in recent years, focusing on creating algorithms that can apply knowledge across different tasks with limited data. MAML has become a key method in this area, providing a flexible system for quickly adapting to new tasks with just a few gradient updates. MAML is designed to grasp a model initialization that can be effectively adjusted for various tasks. This makes it well-suited for robotics, natural language processing, and computer vision. MAML is based on the idea that when training on multiple tasks, the model parameters should be initialized so that they need minimal adjustment to learn a new task. 






    

    


MAML's model-agnostic nature sets it apart from other meta-learning algorithms by not imposing constraints on the underlying model architecture. This flexibility enables MAML to be utilized across a diverse range of learning tasks, including supervised learning and reinforcement learning scenarios. Furthermore, MAML is compatible with various optimization algorithms, allowing it to adapt to different learning environments.

Although MAML offers significant benefits, it also poses challenges, particularly concerning computational efficiency during training. There may be potential instability in the meta-optimization process as well. One of the primary issues is the computational expense associated with the calculation of second-order derivatives during the outer-loop update. To mitigate this, FOMAML approximates the second-order optimization by neglecting the Hessian term, reducing the computational overhead. This simplification results in a more efficient and straightforward update rule, expressed as:

\begin{equation}
    \theta \leftarrow \theta - \alpha \nabla_\theta \mathcal{L}_{T_i}(\theta)
\end{equation}

Here, the symbols are defined as follows:

\begin{itemize}
    \item \( \theta \): The parameter vector of the meta-model, representing the initialization optimized during the meta-training phase.
    \item \( \alpha \): The task-specific learning rate, which controls the magnitude of the parameter update in the inner loop for task adaptation.
    \item \( \mathcal{L}_{T_i}(\theta) \): The loss function associated with task \( T_i \), measuring the discrepancy between the model predictions and the ground truth for the task.
    \item \( \nabla_\theta \mathcal{L}_{T_i}(\theta) \): The gradient of the loss function \( \mathcal{L}_{T_i}(\theta) \) with respect to the parameters \( \theta \), capturing the direction of steepest ascent in the loss landscape.
\end{itemize}

By performing a gradient step along \( \nabla_\theta \mathcal{L}_{T_i}(\theta) \) scaled by \( \alpha \), the model parameters are adapted to minimize the task-specific loss \( \mathcal{L}_{T_i}(\theta) \). This adaptation enables the model to specialize for task \( T_i \) efficiently while preserving the generalizability encoded in \( \theta \).

The computational demands are balanced with this adjustment while maintaining the effectiveness of the meta-learning process. FOMAML, therefore, provides a practical and computationally efficient approach for meta-learning, particularly for applications requiring scalable optimization over multiple tasks.

\subsection{Problem Formulation}

In the context of Metasurface design, our goal is to predict key Metasurface properties, specifically transmittance, reflectance, and absorbance, based on the Metasurface design variables. These design variables include operational parameters such as frequency, capacitance, inductance, resistance, angle of incidence, array size, and inter-element spacing.

Given a limited dataset at lower frequency ranges, we aim to accurately predict Metasurface properties at higher frequencies, effectively extending the model’s predictive range. We formulate this task as a supervised learning problem. Here, we seek a predictive function \( f_{\theta} \) that maps the design variables to the Metasurface properties, with \( \theta \) representing the model parameters optimized through MAML.


Let the Metasurface design variables be represented as an input vector \( \mathbf{x} \in \mathbb{R}^d \), where \( d \) is the number of design features. The corresponding Metasurface properties are captured by the output vector \( \mathbf{y} \in \mathbb{R}^3 \), representing transmittance, reflectance, and absorbance. Thus, the problem is framed as finding a function \( f_{\theta}: \mathbb{R}^d \rightarrow \mathbb{R}^3 \) such that:

\[
f_{\theta}(\mathbf{x}) = \mathbf{y}
\]

To achieve this, we apply a first-order MAML (FOMAML) approach, which enables efficient gradient-based meta-learning by approximating the second-order optimization with a computationally simpler first-order update. Our tasks, sampled randomly from the dataset, are constructed by splitting observations into support and query or target sets, where each task corresponds to predicting Metasurface properties at a higher frequency range based on observed data at a lower frequency range.


Using the FOMAML framework, we aim to identify an optimal initialization for the model parameters \( \theta \) that allows for rapid adaptation to each new task with only a few gradient updates. Given a set of tasks \( T_1, T_2, \ldots, T_K \sim p(T) \), where each task is defined by a support-query split over different frequency ranges, the meta-learning objective is as follows:

\begin{enumerate}
\item \textbf{Task-Specific Loss}: For each task \( T_i \), where observations are sampled from a lower frequency range and predictions are required at a higher frequency range, we define the task-specific loss \( \mathcal{L}_{T_i}(\theta) \) as the mean squared error between the predicted and actual Metasurface properties:

    \[
    \mathcal{L}_{T_i}(\theta) = \frac{1}{N} \sum_{j=1}^{N} \| f_{\theta}(\mathbf{x}_j^{(i)}) - \mathbf{y}_j^{(i)} \|^2
    \]

    where \( \{ (\mathbf{x}_j^{(i)}, \mathbf{y}_j^{(i)}) \}_{j=1}^{N} \) represents the samples for task \( T_i \).

\item \textbf{Inner Update (Task-Specific Adaptation)}: For each task, we perform a gradient update on \( \theta \) to adapt to the specific task \( T_i \). Since we are using FOMAML, the update avoids the computationally expensive second-order gradients, allowing a simplified update as follows:

    \[
    \theta_i' = \theta - \alpha \nabla_{\theta} \mathcal{L}_{T_i}(\theta)
    \]

    where \( \alpha \) is the task-specific learning rate, enabling efficient adaptation to \( T_i \).

\item \textbf{Outer Update (Meta-Objective Optimization)}: The meta-objective is formulated to find an initialization \( \theta \) that minimizes the cumulative adapted loss across tasks, effectively capturing generalizable knowledge across various frequency splits:

    \[
    \min_{\theta} \sum_{i=1}^{K} \mathcal{L}_{T_i}(\theta_i') = \min_{\theta} \sum_{i=1}^{K} \mathcal{L}_{T_i}(\theta - \alpha \nabla_{\theta} \mathcal{L}_{T_i}(\theta))
    \]

This meta-objective seeks to learn a generalizable initialization \( \theta \) that requires minimal adjustment for accurate predictions in new frequency ranges. Hence, it ensures robust performance for Metasurface property prediction across unobserved frequencies.

\end{enumerate}

\begin{figure*}
\centering
\includegraphics[width=0.8\linewidth]{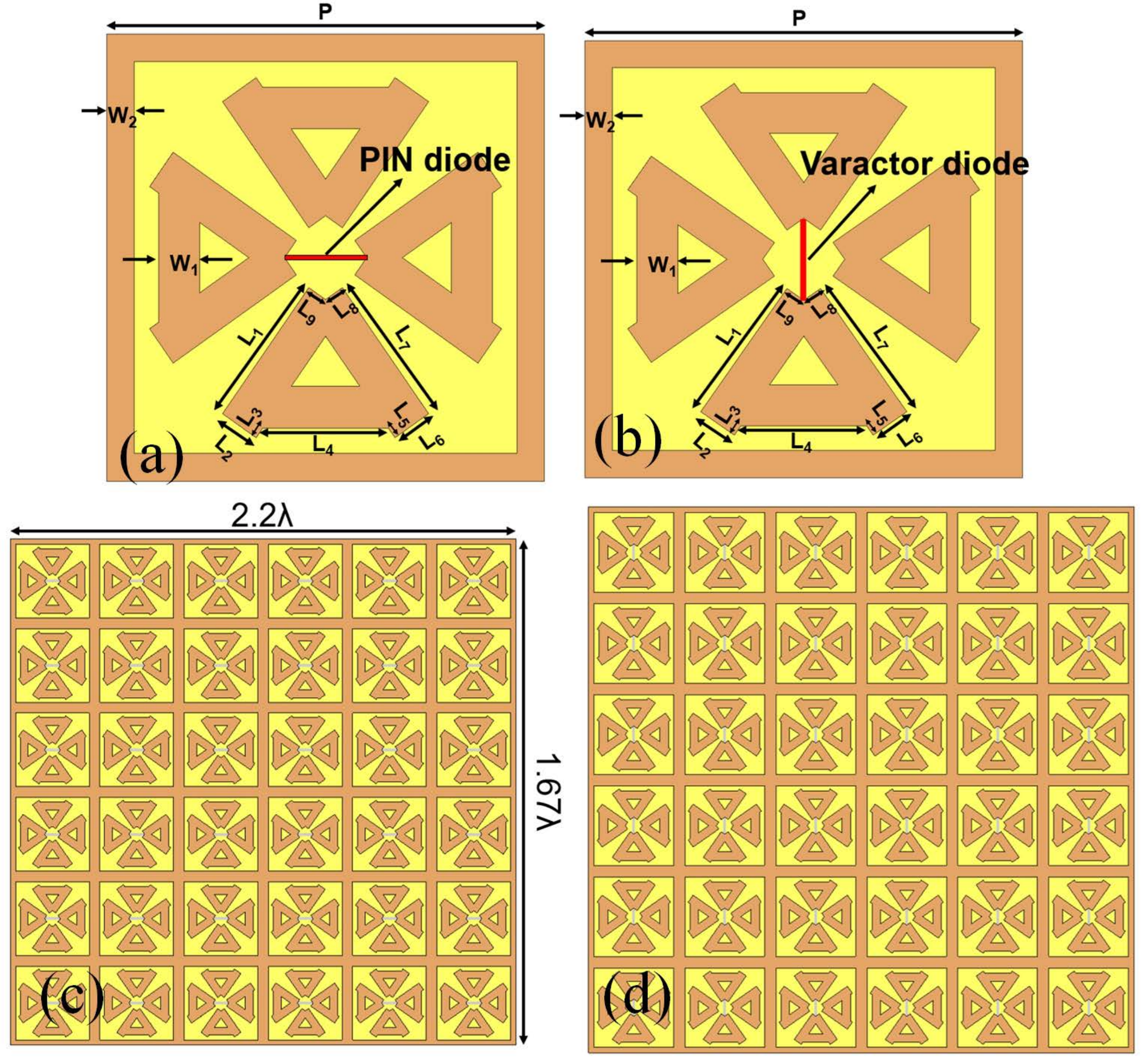}

\caption{(a) The top and (b) the bottom sides of the meta-unit, (c) the top and (d) the bottom view of the metasurface-based $6\times6$ TRA.} 
\label{fig:str1}
\vspace{-0.4 cm}
\end{figure*}

The meta-trained function \( f_{\theta} \) is subsequently fine-tuned on new, higher-frequency data in the meta-test phase, using the initial parameters learned during meta-training. This fine-tuning step is designed to leverage the generalizable knowledge captured through the meta-learning process, enabling accurate predictions with limited higher-frequency data. Thus, the problem is ultimately framed as learning an initialization of \( \theta \) for the function \( f_{\theta}(\mathbf{x}) = \mathbf{y} \) that can generalize efficiently to new frequency domains. This formulation meets the complex requirements of Metasurface property prediction across diverse spectral conditions.

This formulation within the FOMAML framework enables our model to address the variability of frequency-specific Metasurface property prediction while ensuring computational efficiency. It is particularly well-suited for real-world applications where data in the target frequency range may be limited.


\section{Methodology}
In this section, we describe our proposed meta-learning-based framework, MetaFAP. We first outline the dataset generation process, detailing the key variables and features involved. Next, we provide an architectural overview of our proposed base model. Finally, we explain the implementation of our meta-learning framework.

\subsection{Dataset Generation}
The training datasets are generated by managing all potential variable features of a metasurface-based transmit reflect array (TRA). Fig.~\ref{fig:str1}(a) and Fig.~\ref{fig:str1}(b) depict the upper and lower surfaces, respectively, of the proposed meta-unit. The proposed meta-unit consists of two patches separated by a dielectric layer. Four identical triangular-shaped structures are slotted in the top and bottom patches. A PIN diode and a varactor are strategically incorporated between the horizontal and vertical triangular structures on the top and bottom metallic layers, respectively. The dielectric layer consists of FR-4, demonstrating a relative permittivity of 4.3 and a loss tangent of 0.025, with a thickness of 2 mm. The meta-unit exhibits a periodicity P of 0.22$\lambda$, with the wavelength $\lambda$ measuring 45 mm. The reconfigurability of the meta-unit, including transmittance and reflectance, is enabled through the PIN diode and the varactor into the upper and lower layers, respectively. The geometric dimensions of the design are specified in detail as; $L_1$ = 5.5 mm, $L_2$ = 1.5 mm,  $L_3$ = 0.4 mm, $L_4$ = 4.6 mm, $L_5$ = 0.4 mm, $L_6$ = 1.5 mm, $L_7$ = 5.5 mm, $L_8$ = 0.75 mm, $L_9$ = 0.75 mm, $W_1$ = 1.5 mm, and $W_2$ = 1 mm. The MACOM MA46H120 PIN diode is utilized as a switchable element, modeled through a series of lumped parameters: $R_v$, $L_v$, and $C_{vt}$ \cite{Dai2020_Access}. The varactor diode is Skyworks SMV1247 exhibiting a tuning capacitance, $C_{vb}$ within the range of 0.64 to 8.86 pF. A 6 × 6 (2.2$\lambda$ × 1.67$\lambda$) $mm^2$ TRA is devised. The influence of eight design features, including frequency varying from 5 to 11 GHz, and 15 to 25 GHz, incident angle ($\theta$) ranging from 0 to 90$^{\circ}$, spacing between unit cells ($d$) varying from $\lambda$/2 to $\lambda$/4, $C_{vt}$ ranging from 50 to 353 fF, $C_{vb}$ ranging from 0.64 to 8.86 pF, $R_v$ ranging from 0.8 to 50 $\Omega$, $L_v$ ranging from 750 to 850 pH, array size ($A$) ranging from 2 × 2 to 6 × 6, etc., on the transmittance and reflectance are simulated. Although full-wave EM simulations with adequate mesh density provide reliable evaluation results, they come with considerable computational latency. To address this, we develop MetaFAP to efficiently predict transmittance, reflectance, and absorbance across a broad frequency range by learning the behavior of the TRA from the simulated data.

\subsection{Architectural Overview of the Base Model}

\begin{figure*}[htp]
    \centering
    \includegraphics[width=1\textwidth, trim={12.8cm 4.9cm 12.8cm 4.9cm}, clip]{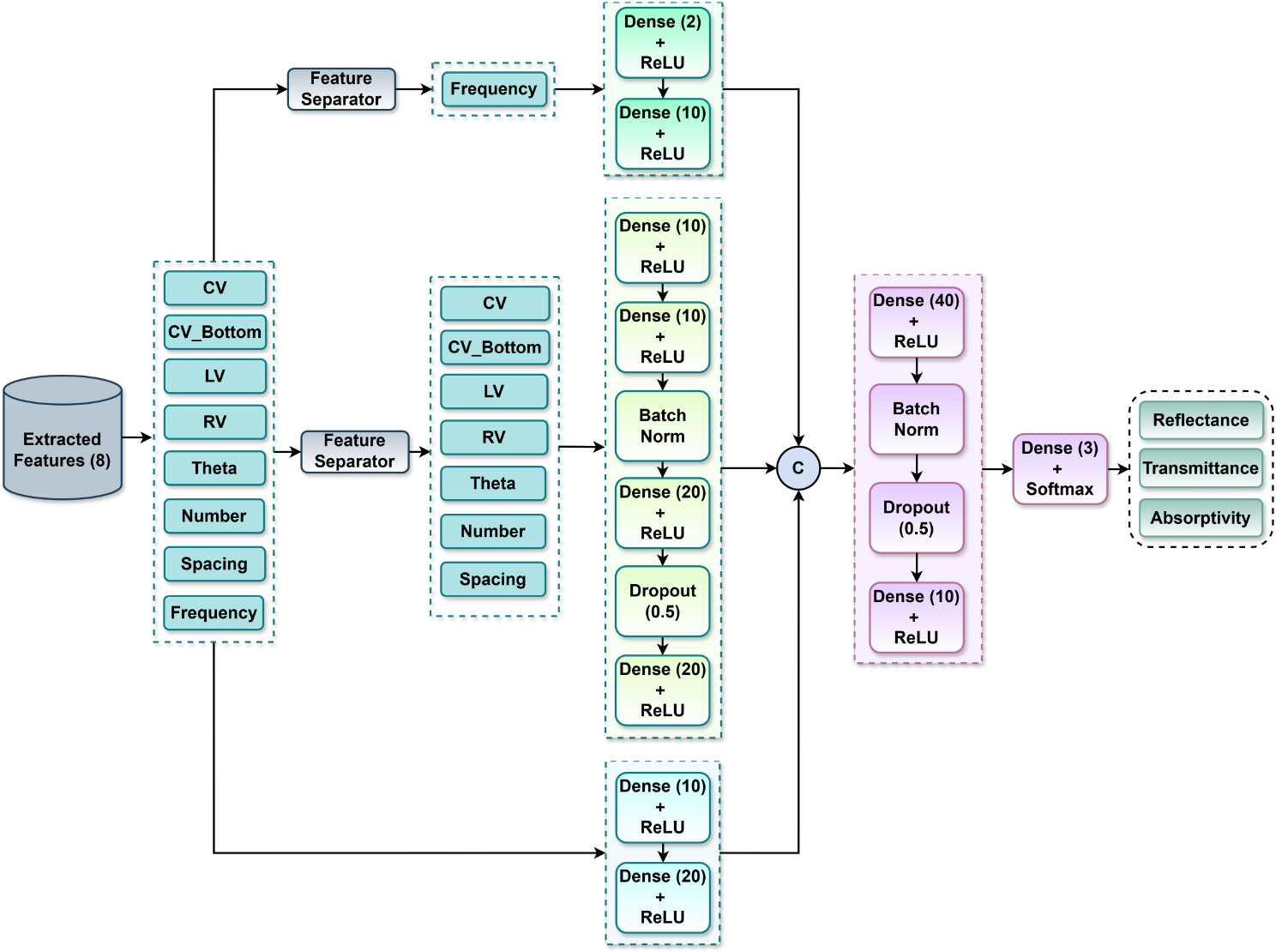}
    \caption{Proposed MetaFAP Base-model Architecture: The model splits the input into frequency and non-frequency features, processes each branch through specialized layers, and then concatenates them for final prediction of reflectance, transmittance, and absorptivity.}
    \label{fig:base_model}
\end{figure*}

The proposed base model is designed to effectively predict metasurface properties by leveraging a modular architecture that integrates specialized networks for different feature subsets. This hybrid structure incorporates distinct pathways for frequency-related and non-frequency-related features, ensuring accurate predictions across diverse operating conditions. The base-model architecture is displayed in Figure \ref{fig:base_model}.

\subsubsection{Input Representation and Feature Segmentation}

The input to the model comprises an 8-dimensional feature vector representing metasurface design parameters, including frequency, capacitance, inductance, resistance, angle of incidence, unit cell count, and inter-element spacing. These features are systematically partitioned into two subsets: (i) frequency-dependent features and (ii) non-frequency-related design parameters. Mathematically, let the input vector be defined as:
\[
\mathbf{x} = [x_f, \mathbf{x}_o] \in \mathbb{R}^8,
\]
where \( x_f \) represents the frequency-related scalar input, and \( \mathbf{x}_o \) denotes the remaining design variables.

\subsubsection{Gating Mechanism for Global Feature Interactions}

To capture global dependencies among all input features, a gating network is employed. This network operates on the entire input vector \(\mathbf{x}\) and generates adaptive gating signals to regulate feature importance. The gating operation is expressed as:
\[
\mathbf{h}_g = g(\mathbf{x}; \Theta_g),
\]
where \(g(\cdot)\) represents a series of nonlinear transformations parameterized by \(\Theta_g\), including fully connected layers with activation functions and regularization techniques.

\subsubsection{Independent Network for Non-Frequency Features}

The non-frequency features are processed through a dedicated network optimized for extracting complex interdependencies within these parameters. This pathway consists of multiple dense layers interspersed with normalization and dropout layers to enhance stability and mitigate overfitting. Let \(\mathbf{h}_o\) denote the extracted features, computed as:
\[
\mathbf{h}_o = f_o(\mathbf{x}_o; \Theta_o),
\]
where \(f_o(\cdot)\) is the mapping function of the independent network.

\subsubsection{Frequency-Specific Network}

The frequency pathway is tailored to capture nuanced dependencies associated with frequency-specific variations in metasurface properties. This network comprises lightweight dense layers to ensure efficient computation while preserving critical information. The frequency-related features are computed as:
\[
\mathbf{h}_f = f_f(x_f; \Theta_f),
\]
where \(f_f(\cdot)\) represents the frequency-specific transformation.

\subsubsection{Feature Integration and Prediction}

The outputs from the gating, non-frequency, and frequency-specific networks (\( \mathbf{h}_g, \mathbf{h}_o, \mathbf{h}_f \)) are concatenated to form a unified feature representation:
\[
\mathbf{h}_c = [\mathbf{h}_g, \mathbf{h}_o, \mathbf{h}_f].
\]
This integrated representation is further processed through fully connected layers, incorporating dropout and normalization to enhance generalization. The final output layer employs a softmax function to predict metasurface properties, represented as:
\[
\hat{\mathbf{y}} = \text{softmax}(\mathbf{W}_o \mathbf{h}_c + \mathbf{b}_o),
\]
where \(\hat{\mathbf{y}} \in \mathbb{R}^3\) corresponds to the predicted probabilities for transmittance, reflectance, and absorbance.

The modular design of the base model ensures robust learning of both global and local feature interactions, enabling precise metasurface property prediction across a wide range of design configurations.

\subsection{MAML Framework}

To tackle the challenge of predicting Metasurface properties across a broad frequency range, we adopt MAML as the underlying meta-learning approach. MAML is particularly suited for tasks requiring rapid adaptation to new but related tasks with limited data. In our scenario, predicting Metasurface properties such as transmittance, reflectance, and absorbance at higher frequencies (22-25 GHz) is challenging due to the limited training data availability at these frequencies. Therefore, we train our model primarily on low-frequency data (5-11 GHz) and leverage MAML to enable efficient adaptation to higher frequencies during inference.

\subsubsection{Data Preparation for Meta-Learning}

The dataset contains design variables such as frequency, capacitance, inductance, resistance, angle of incidence, array size (number of unit cells), and inter-element spacing as features to predict Metasurface properties. To apply MAML effectively, we split the data into three sets: meta-train, meta-validation, and meta-test.

\begin{itemize}
    \item \textbf{Meta-Train Set:} This set includes data with a frequency range of 5-11 GHz and 15-16.5 GHz as the support set and 16.5-19 GHz as the query set. This division allows the model to learn patterns from low-frequency data while adapting to a slightly higher frequency range.
    \item \textbf{Meta-Validation and Meta-Test Sets:} Both sets use frequencies of 19-22 GHz as support and 22-25 GHz as query. These sets are critical for assessing the model’s ability to generalize and adapt to new frequency ranges, thereby evaluating the effectiveness of the MAML approach.
\end{itemize}

Each data subset is further processed to ensure consistent scaling across features. We apply standard scaling to the meta-train data and reuse this scaling transformation across the meta-validation and meta-test sets, maintaining consistency in feature representation across different splits.

\subsubsection{MAML Training Framework}

MAML optimizes for a set of model parameters that serve as a good initialization for learning new tasks. The MAML training process consists of inner-loop and outer-loop updates:

\begin{itemize}
    \item \textbf{Inner Loop (Task-Specific Adaptation):} For each task, MAML performs a few gradient updates using the support set, fine-tuning the model parameters to minimize the loss on this support data. This adaptation allows the model to specialize in each task's specific requirements. In our implementation, the inner loop employs the Stochastic Gradient Descent (SGD) optimizer with a learning rate of $\beta = 0.0005$ for task-specific updates, ensuring controlled and smooth adaptation.

    \item \textbf{Outer Loop (Meta-Update):} After adapting to each task, the model's performance is evaluated on the query or target set. The gradients calculated from the query set are then averaged across all tasks and used to update the initial model parameters. This meta-update, managed by the AdaBelief optimizer with a learning rate $\alpha = 0.0003$, aims to create a robust initialization that enables rapid adaptation to new tasks. Learning rates for both inner and outer loops are dynamically adjusted during training to ensure optimal convergence.
\end{itemize}

\begin{figure*}[htp]
    \centering
    \includegraphics[width=1\textwidth, trim={0.8cm 1.5cm 0.8cm 1.5cm}, clip]{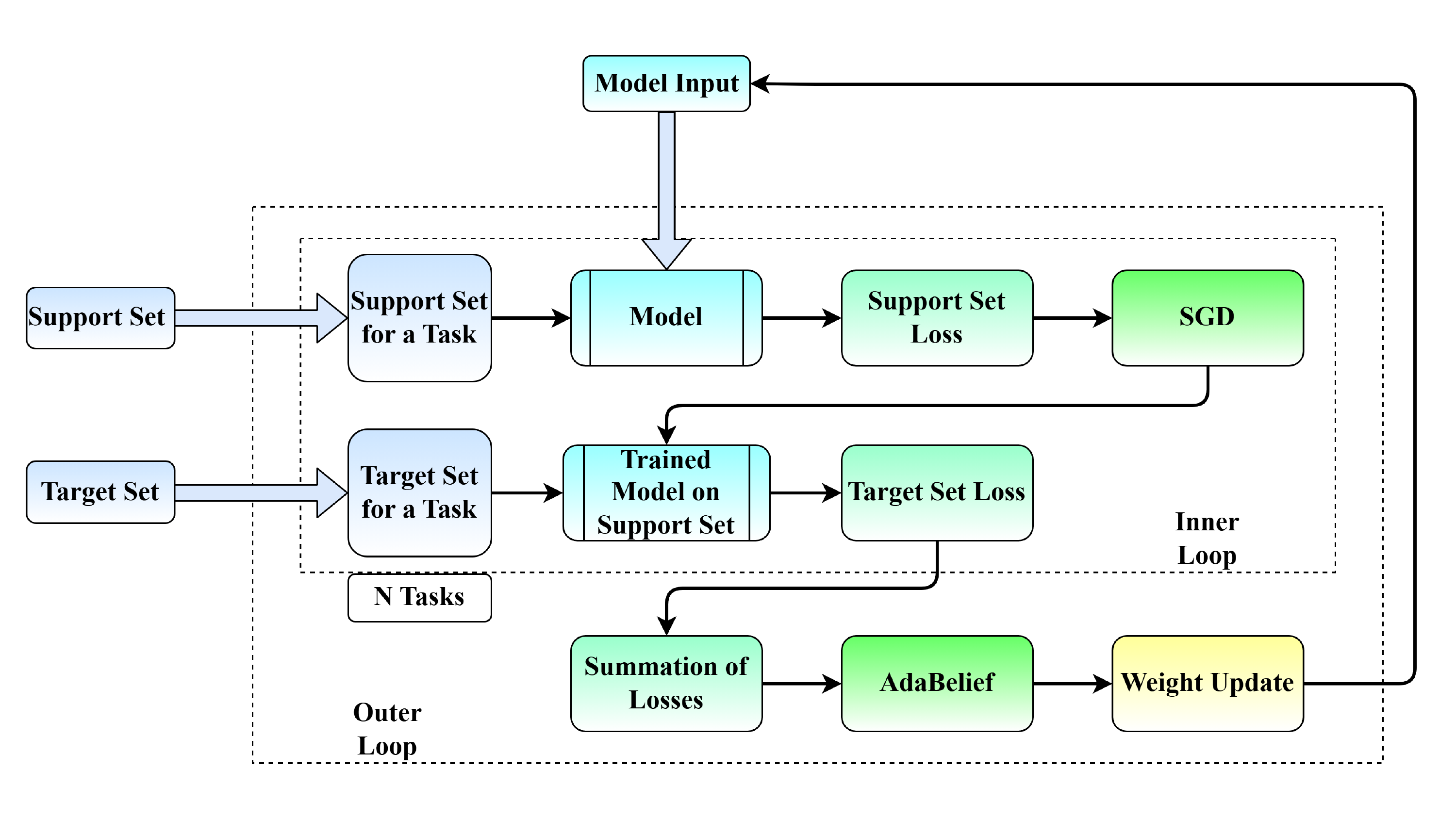}
    \caption{MetaFAP-Learning Process: The model is first adapted to each task using support-set data and an inner-loop optimizer (SGD). After evaluating on the corresponding target set, losses from multiple tasks are aggregated in the outer loop and minimized via AdaBelief, updating the model’s initial parameters for improved few-shot performance across diverse tasks.}
    \label{fig:maml_framework}
\end{figure*}

The entire MetaFAP learning process is depicted in Figure \ref{fig:maml_framework}. The following key components facilitate this MAML framework:

\begin{itemize}
    \item \textbf{Meta-Gradient Accumulation:} For each task, gradients obtained from the query set are accumulated, and their average is used to update the model’s initial parameters. This approach captures generalizable knowledge that benefits the model's adaptation to unseen frequencies.
  
    \item \textbf{Adaptive Learning Rate Scheduling:} Both task-specific and meta-update learning rates are scheduled based on the epoch, decreasing over time to stabilize training and prevent overshooting. This dynamic learning rate adjustment enhances the model’s stability, improving the generalization of Metasurface property predictions across various frequency ranges.
\end{itemize}

\subsubsection{Loss Functions and Evaluation Metrics}
The model is trained via the Hubcor loss function \cite{murshed2024real}, which has demonstrated superior performance compared to alternative regression losses \cite{murshed2023cnn}.
To evaluate the model’s predictions on Metasurface properties, we use Mean Squared Error (MSE) as the primary metric. In addition, we compute the Pearson correlation coefficient between predicted and actual values to measure the alignment in prediction trends. These metrics are calculated for both the support and query sets within each task, providing insight into the model's performance during task-specific adaptation and generalization across tasks.

The support set performance reflects how well the model adapts to the provided low-frequency data, while the query set performance indicates its generalization ability to the targeted higher frequency ranges. During training, the model's weights are saved whenever the validation loss improves, preserving the best model configuration for prediction on new data.

\subsubsection{Evaluation and Prediction on Meta-Test Set}

After training, we evaluate the model on the meta-test set, which is adapted using the support data and subsequently tested on the query data. For each query batch, the model’s initial parameters are adapted to the support set of the task using a few gradient steps, followed by predictions on the query set. The task-specific adaptation and evaluation ensure the model can generalize to higher frequencies, leveraging the initial parameters learned from the meta-training process.

In summary, the MetaFAP framework equips the model with a flexible initialization, enabling rapid adaptation to new tasks in the target frequency range. By learning to adapt to frequency shifts within the Metasurface data, our approach successfully addresses the challenge of extrapolating Metasurface properties from low-frequency training data to higher-frequency prediction tasks. This meta-learning strategy is therefore a suitable choice for our task, as it allows the model to generalize efficiently to unseen frequency domains with limited support data. This makes it an optimal solution for frequency-variant Metasurface property prediction.

\subsection{Model Testing and Evaluation}

To evaluate the effectiveness of MetaFAP for predicting Metasurface properties, we conducted experiments using a structured meta-test dataset. This enables a comprehensive assessment of the model’s capacity to generalize and adapt across a broad frequency spectrum. Here, we outline the testing procedures and performance metrics employed, followed by an analysis of the evaluation results.

\begin{figure}[!htbp]
\centering
\includegraphics[width=\columnwidth]{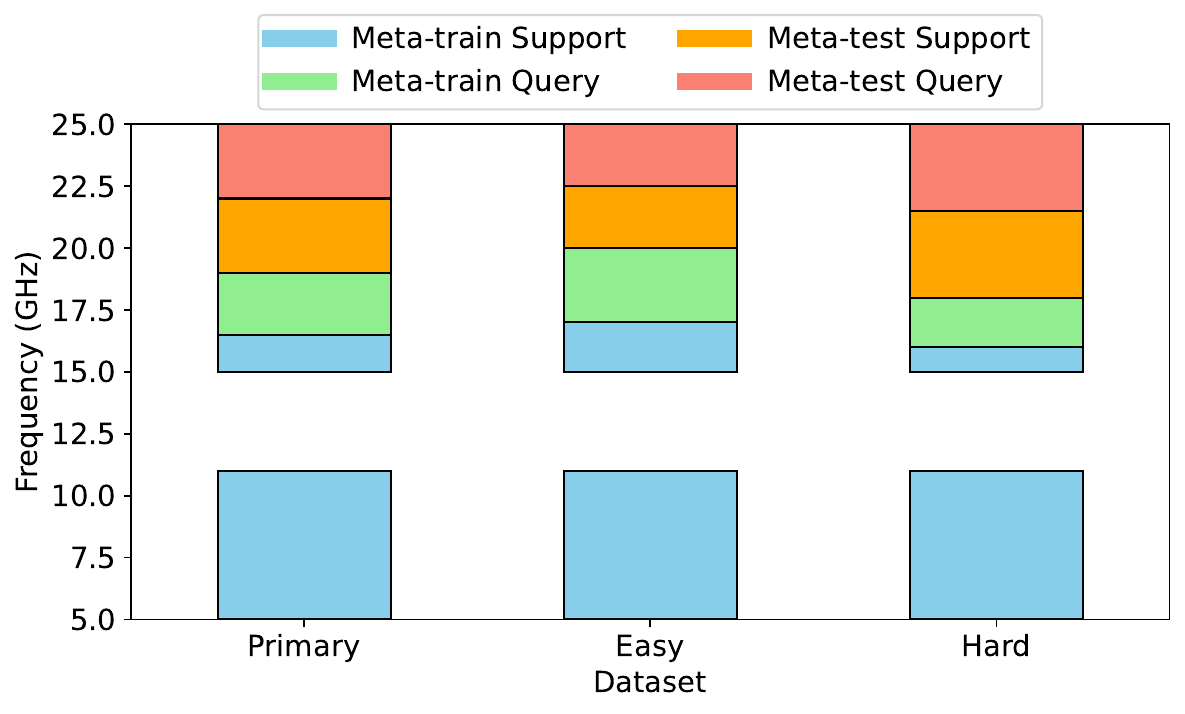}
\caption{A plot showing the distribution of data in the three datasets of varying difficulty.}
\label{fig:dataset}
\end{figure}

\subsubsection{Experimental Setup}

Following the meta-training phase, we applied the trained model to a meta-test set comprising frequencies outside the model's training range. We provide our model with three types of datasets: primary, easy, and hard. In the primary dataset, the meta-train set contains support data in the range of 5–11 GHz and 15–16.5 GHz, with corresponding query data in the range of 16.5–19 GHz. Meanwhile, the meta-train and meta-test sets for this dataset include support data in the range of 19–22 GHz and query data in the range of 22–25 GHz, presenting a moderate challenge to the model. The easy dataset is constructed with a meta-train set having support data in the range of 5–11 GHz and 15–17 GHz, and query data in the range of 17–20 GHz. The meta-train and meta-test sets for this `easy dataset' feature support data in the range of 20–22.5 GHz and query data in the range of 22.5–25 GHz, thereby presenting an easier challenge. The hardest dataset is designed with a meta-train set that includes support data in the range of 5–11 GHz and 15–16 GHz, and query data in the range of 16–18 GHz; here, the meta-train and meta-test sets incorporate support data in the range of 18–21.5 GHz and query data in the range of 21.5–25 GHz, representing the toughest challenge for the model. The complete dataset distribution is illustrated in Figure \ref{fig:dataset}. During evaluation, the model’s weights were first initialized with the optimized parameters learned during meta-training. Subsequently, the model adapted to the meta-test support set, and its performance was evaluated on the meta-test query set to measure its generalization and adaptability across frequency shifts.

\subsubsection{Evaluation Metrics}

Performance was quantified through three key metrics to ensure a comprehensive assessment of the model's predictive accuracy:

\begin{itemize}
    \item \textbf{Mean Squared Error (MSE):} This metric measures the average squared deviation between predicted and actual Metasurface property values, offering insight into the model's overall prediction accuracy across varying frequency data.
    \item \textbf{Mean Absolute Error (MAE):} MAE quantifies the average absolute error, providing an intuitive measure of the model’s accuracy in absolute terms, which is beneficial for understanding real-world deviations in Metasurface properties.
    \item \textbf{Pearson Correlation Coefficient (R-Score):} This coefficient evaluates the linear correlation between the actual and predicted values, highlighting the model's ability to maintain trend accuracy across predictions.
\end{itemize}


\section{Results}

This section provides a comprehensive evaluation of MetaFAP's performance against conventional ML and DL benchmarks. Initially, we present a comparative analysis on three distinct datasets, highlighting quantitative improvements across multiple performance metrics. Subsequently, we assess the model's robustness under varying data conditions and conduct an extensive ablation study to pinpoint the key contributors to performance.

\subsection{Comparison with other Models on three different datasets}

Table \ref{tab:comparison} summarizes the performance of various models across three distinct dataset distributions—Primary, Easy, and Hard—using MSE, MAE, and CC as evaluation metrics. Traditional regression models (e.g., DecisionTreeRegressor, RandomForestRegressor, KNeighborsRegressor, BaggingRegressor, and ExtraTreesRegressor) generally yield moderate performance, with relatively high MSE and MAE values on the Primary and Hard datasets and only modest CCs (around 6–15\%). Their performance improves on the Easy dataset, achieving CC values in the mid-60s, which suggests that these methods can capture the underlying trends when the data distribution is simpler or less noisy. However, their inability to model complex nonlinear interactions and adapt to distribution shifts limits their effectiveness in more challenging scenarios.

Boosting-based models, including GradientBoostingRegressor, XGBRegressor, LGBMRegressor, and HistGradientBoosting, also struggle with generalization; some even produce negative correlation coefficients on the Primary dataset. This outcome indicates that the aggregation of weak learners in these methods is insufficient to robustly capture the intricate behavior of metasurface properties, particularly when confronted with out-of-distribution data. In stark contrast, the traditional deep learning model, IR-DNN \cite{IR_DNN}, exhibits severe overfitting, as evidenced by extremely high MSE and MAE values along with very negative CC percentages. Such behavior is symptomatic of models that have memorized the training data distribution but fail to generalize when faced with novel or shifted conditions.

\begin{figure*}[h]
    \centering
    \includegraphics[width=1\textwidth, trim={0.1cm 0.3cm 0.3cm 0.1cm}, clip]{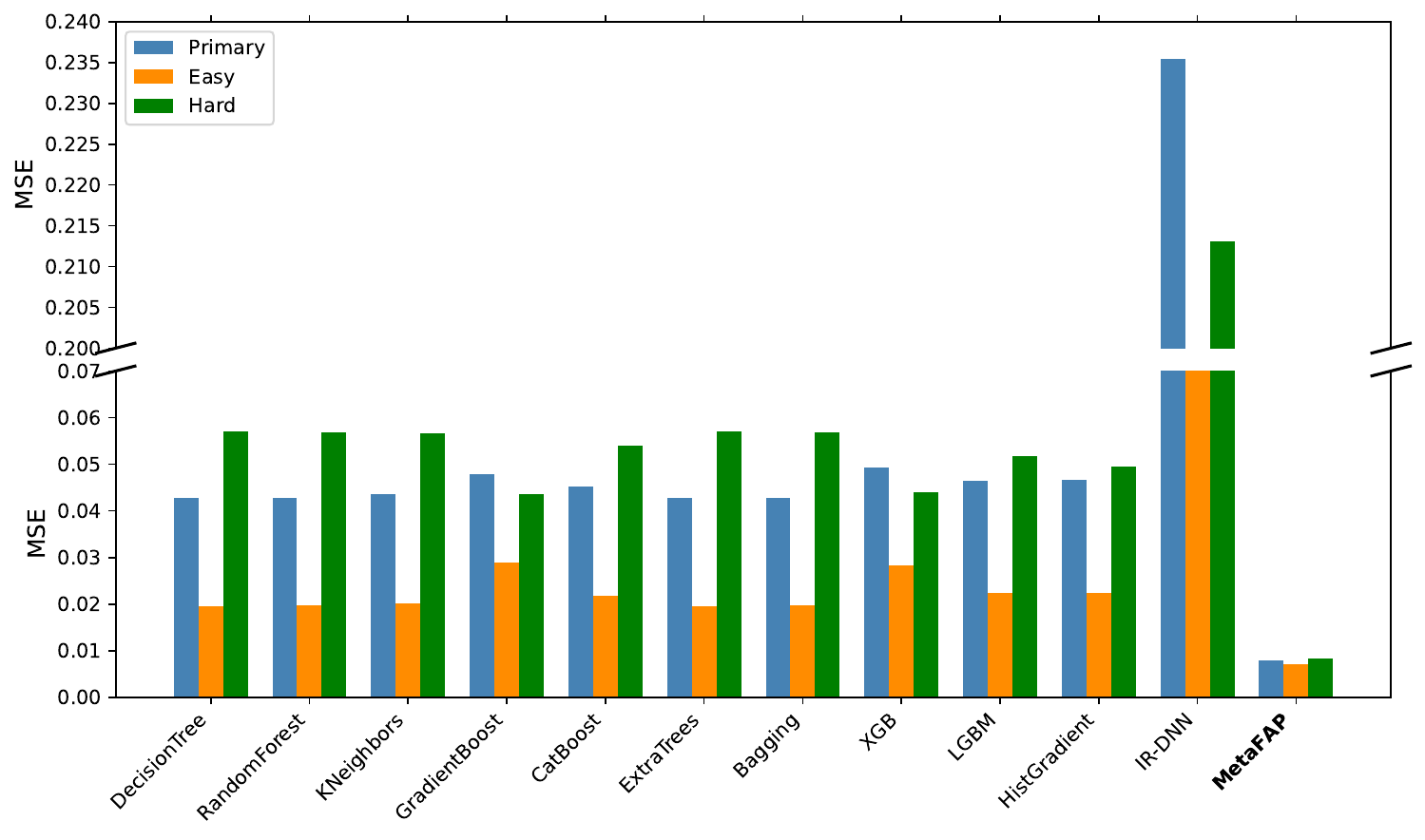}
    \caption{Bar plot presenting the performance comparison according to MSE of the proposed MetaFAP with the traditional models on primary, easy, and hard data split.}
    \label{fig:bar_plot}
\end{figure*}

Our proposed model, MetaFAP, outperforms all competitors with an order-of-magnitude improvement in MSE and MAE, and achieves CC values above 77\% across all datasets. This superior performance can be attributed to the meta-learning framework underlying MetaFAP, which learns a robust initialization that enables rapid adaptation to new tasks with minimal additional data. As a result, MetaFAP is highly resilient to distribution shifts, effectively handling the complexities of cross-frequency and heterogeneous metasurface data. In contrast, conventional DL models like IR-DNN lack such adaptive capabilities, leading to overfitting and poor performance when encountering out-of-distribution samples. Figure \ref{fig:bar_plot} clearly depicts the superiority of MetaFAP in all three data sets over the other models.

\begin{table}[h]
\centering
\caption{Performance Comparsion with other Models on three different datasets}
\begin{tabular}{|p{0.35\linewidth}|c|c|c|c|}
\hline
\textbf{Model} & Dataset & \textbf{MSE} & \textbf{MAE} & \textbf{CC \%} \\ \hline
    \multirow{3}{*}{DecisionTreeRegressor} & Primary & 0.0427 & 0.1718 & 6.11 \\
    \cline{2-5}
    & Easy & 0.0196 & 0.1108 & 65.33 \\
    \cline{2-5}
    & Hard & 0.0570 & 0.1942 & 14.36 \\
    \hline

    \multirow{3}{*}{RandomForestRegressor} & Primary & 0.0427 & 0.1719 & 6.06 \\
    \cline{2-5}
    & Easy & 0.0198 & 0.1116 & 64.79 \\
    \cline{2-5}
    & Hard & 0.0568 & 0.1938 & 14.36 \\
    \hline

        \multirow{3}{*}{KNeighborsRegressor} & Primary & 0.0435 & 0.1743 & 4.75 \\
    \cline{2-5}
    & Easy & 0.0201 & 0.1127 & 63.96 \\
    \cline{2-5}
    & Hard & 0.0566 & 0.1938 & 14.97 \\
    \hline

        \multirow{3}{*}{GradientBoostingRegressor} & Primary & 0.0479 & 0.1928 & -7.54 \\
    \cline{2-5}
    & Easy & 0.0289 & 0.1423 & 38.63 \\
    \cline{2-5}
    & Hard & 0.0435 & 0.1731 & 29.07 \\
    \hline

        \multirow{3}{*}{CatBoostRegressor} & Primary & 0.0452 & 0.1804 & 0.41 \\
    \cline{2-5}
    & Easy & 0.0218 & 0.1185 & 59.36 \\
    \cline{2-5}
    & Hard & 0.0539 & 0.1891 & 16.34 \\
    \hline

        \multirow{3}{*}{ExtraTreesRegressor} & Primary & 0.0427 & 0.1718 & 6.11 \\
    \cline{2-5}
    & Easy & 0.0196 & 0.1108 & 65.34 \\
    \cline{2-5}
    & Hard & 0.0570 & 0.1942 & 13.77 \\
    \hline

        \multirow{3}{*}{BaggingRegressor} & Primary & 0.0427 & 0.1719 & 6.06 \\
    \cline{2-5}
    & Easy & 0.0198 & 0.1116 & 64.79 \\
    \cline{2-5}
    & Hard & 0.0568 & 0.1938 & 14.37 \\
    \hline

        \multirow{3}{*}{XGBRegressor} & Primary & 0.0492 & 0.1956 & -9.35 \\
    \cline{2-5}
    & Easy & 0.0283 & 0.1416 & 39.48 \\
    \cline{2-5}
    & Hard & 0.0439 & 0.1718 & 29.94 \\
    \hline

        \multirow{3}{*}{LGBMRegressor} & Primary & 0.0465 & 0.1846 & -2.06 \\
    \cline{2-5}
    & Easy & 0.0224 & 0.1212 & 57.55 \\
    \cline{2-5}
    & Hard & 0.0517 & 0.1872 & 17.26 \\
    \hline

        \multirow{3}{*}{HistGradientBoosting} & Primary & 0.0466 & 0.1849 & -2.03 \\
    \cline{2-5}
    & Easy & 0.0223 & 0.1211 & 57.77 \\
    \cline{2-5}
    & Hard & 0.0496 & 0.1835 & 19.65 \\
    \hline

        \multirow{3}{*}{IR-DNN \cite{IR_DNN}} & Primary & 0.2355 & 0.4504 & -78.67 \\
    \cline{2-5}
    & Easy & 0.1032 & 0.2907 & -38.19 \\
    \cline{2-5}
    & Hard & 0.2131 & 0.4333 & -59.23 \\
    \hline

    \multirow{3}{*}{\textbf{MetaFAP}} & \textbf{Primary} & \textbf{0.0079} & \textbf{0.0612} & \textbf{78.54} \\
    \cline{2-5}
    & \textbf{Easy} & \textbf{0.0072} & \textbf{0.0559} & \textbf{81.33} \\
    \cline{2-5}
    & \textbf{Hard} & \textbf{0.0084} & \textbf{0.0732} & \textbf{77.99} \\
    \hline
    
\end{tabular}
\label{tab:comparison}
\end{table}

\subsection{Performance with Varying Tasks and Input Samples}
In this study, we evaluate the effect of varying the number of input support samples on the meta-test performance of our Metasurface property prediction model. Specifically, we assess the model's performance by altering the number of support samples available during the adaptation phase as well as by varying the number of tasks utilized in the meta-training phase. This approach enables a comprehensive analysis of the model's adaptability and stability across different data availability conditions. It is particularly critical for applications where data scarcity may limit prediction accuracy.

\subsubsection{Impact of Varying Support Samples in Meta-Test}
To understand the model's behavior under different sample constraints during the meta-test phase, we conducted experiments by providing 64, 128, 512, and 1024 support samples for adaptation. Table \ref{tab:support_samples} presents the performance metrics—MSE, MAE, and CC—for each sample size configuration.

As shown in Table \ref{tab:support_samples}, increasing the number of support samples generally improves performance. The model achieves its best performance with 1024 samples, yielding an MSE of 0.0078, an MAE of 0.0603, and a CC of 78.53\%. This trend demonstrates the model's capability to leverage a larger volume of support samples for more accurate predictions, as increased data allows for a finer tuning of model parameters during adaptation. However, even with only 64 samples, the model maintains reasonably good performance, indicating its robustness in sample-limited scenarios, which is essential for practical deployment where data acquisition may be costly or restricted.

\begin{table}[h]
\centering
\caption{Performance with Varying Support Samples during Meta-Test}
\begin{tabular}{|c|c|c|c|}
\hline
\textbf{Number of Samples} & \textbf{MSE} & \textbf{MAE} & \textbf{CC \%} \\ \hline
64 & 0.0092 & 0.0782 & 71.05 \\ \hline
128 & 0.0087 & 0.0742 & 74.21 \\ \hline
512 & 0.0079 & 0.0612 & 78.54 \\ \hline
1024 & 0.0078 & 0.0603 & 78.53 \\ \hline
\end{tabular}
\label{tab:support_samples}
\end{table}

\subsubsection{Impact of Varying the Number of Tasks during Meta-Training}
In addition to varying support samples, we explored the model's sensitivity to the number of tasks included during meta-training, experimenting with 32, 64, 128, and 256 tasks. This variation assesses the impact of task diversity on the model’s meta-initialization, which is essential for rapid adaptation across different scenarios. The results, as presented in Table \ref{tab:task_number}, reveal that increasing the number of tasks up to 128 improves model performance, with an optimal MSE of 0.0076, MAE of 0.0602, and CC of 79.45\%. This outcome suggests that higher task diversity during meta-training enhances the model's generalization capabilities, allowing it to adapt more effectively to novel tasks. However, a further increase to 256 tasks shows a slight performance drop, possibly due to increased task variability that may introduce conflicting gradients, thereby complicating the meta-update process.

\begin{table}[h]
\centering
\caption{Performance with Varying Number of Tasks during Meta-Training}
\begin{tabular}{|c|c|c|c|}
\hline
\textbf{Number of Tasks} & \textbf{MSE} & \textbf{MAE} & \textbf{CC \%} \\ \hline
32 & 0.0092 & 0.0785 & 72.89 \\ \hline
64 & 0.0079 & 0.0612 & 78.54 \\ \hline
128 & 0.0076 & 0.0602 & 79.45 \\ \hline
256 & 0.0081 & 0.0633 & 78.29 \\ \hline
\end{tabular}
\label{tab:task_number}
\end{table}

These experiments underscore the model's robustness and flexibility in adapting to various data conditions. They highlight our model's applicability for real-world Metasurface-based communication systems where sample availability and task diversity can vary. The results demonstrate that while increasing both support samples and tasks improves model performance, there exists a certain balance. Sometimes, the addition of too many tasks may counteract the gains achieved through broader task representation.

\subsection{Ablation Study}
To understand the contributions of each architectural component in our base model, we conducted an ablation study by selectively removing different branches of the network and evaluating the impact on performance. Specifically, we analyzed the effect of removing the frequency branch and the non-frequency (other-features) branch on the overall model performance. The results, summarized in Table \ref{tab:ablation_study}, provide insights into the significance of each component in predicting Metasurface properties accurately.

\begin{table}[h]
\centering
\caption{Performance Metrics with Different Model Architectures}
\begin{tabular}{|c|c|c|c|}
\hline
\textbf{Architecture} & \textbf{MSE} & \textbf{MAE} & \textbf{CC \%} \\ \hline
Complete Model & 0.0079 & 0.0612 & 78.54 \\ \hline
No Frequency Branch & 0.0097 & 0.0772 & 72.73 \\ \hline
No Other Features Branch & 0.0082 & 0.0719 & 78.01 \\ \hline
\end{tabular}
\label{tab:ablation_study}
\end{table}

\subsubsection{Complete Model}
The complete model, which includes both the frequency and non-frequency branches, achieved the best performance with an MSE of 0.0079, an MAE of 0.0612, and a CC of 78.54\%. This configuration integrates both frequency-dependent and independent features, enabling the model to capture a comprehensive representation of the Metasurface properties. These results serve as a baseline for comparison with the modified architectures.

\subsubsection{Impact of Removing the Frequency Branch}
When the frequency branch was removed, the model's performance degraded significantly, resulting in an MSE of 0.0097, an MAE of 0.0772, and a CC of 72.73\%. The frequency branch is crucial for processing frequency-specific information, which plays a pivotal role in accurately predicting Metasurface properties across varying frequency ranges. The performance drop highlights the importance of this branch in effectively modeling frequency-dependent variations.

\subsubsection{Impact of Removing the Other-Features Branch}
In the absence of the non-frequency features branch, the model showed a moderate decline in performance, with an MSE of 0.0082, an MAE of 0.0719, and a CC of 78.01\%. Although this degradation is less severe compared to the removal of the frequency branch, it indicates that non-frequency features (such as capacitance, inductance, and resistance) contribute valuable information that enhances the model’s predictive accuracy. The slight reduction in CC also suggests that the model benefits from these features in maintaining strong correlation trends with the ground truth.

\subsubsection{Analysis and Insights}
The ablation study confirms that both branches are essential for optimal performance, with the frequency branch being particularly crucial for the accurate prediction of Metasurface properties. While the non-frequency branch is also important, the model demonstrates a degree of robustness, as the performance decline is less drastic when this branch is omitted. These findings underscore the necessity of a hybrid architecture that leverages both frequency-dependent and independent features, affirming the design choice of the base model as appropriate for the complex task of Metasurface property prediction across diverse frequency ranges.

\begin{figure}[!htbp]
\centering
\includegraphics[width=\columnwidth]{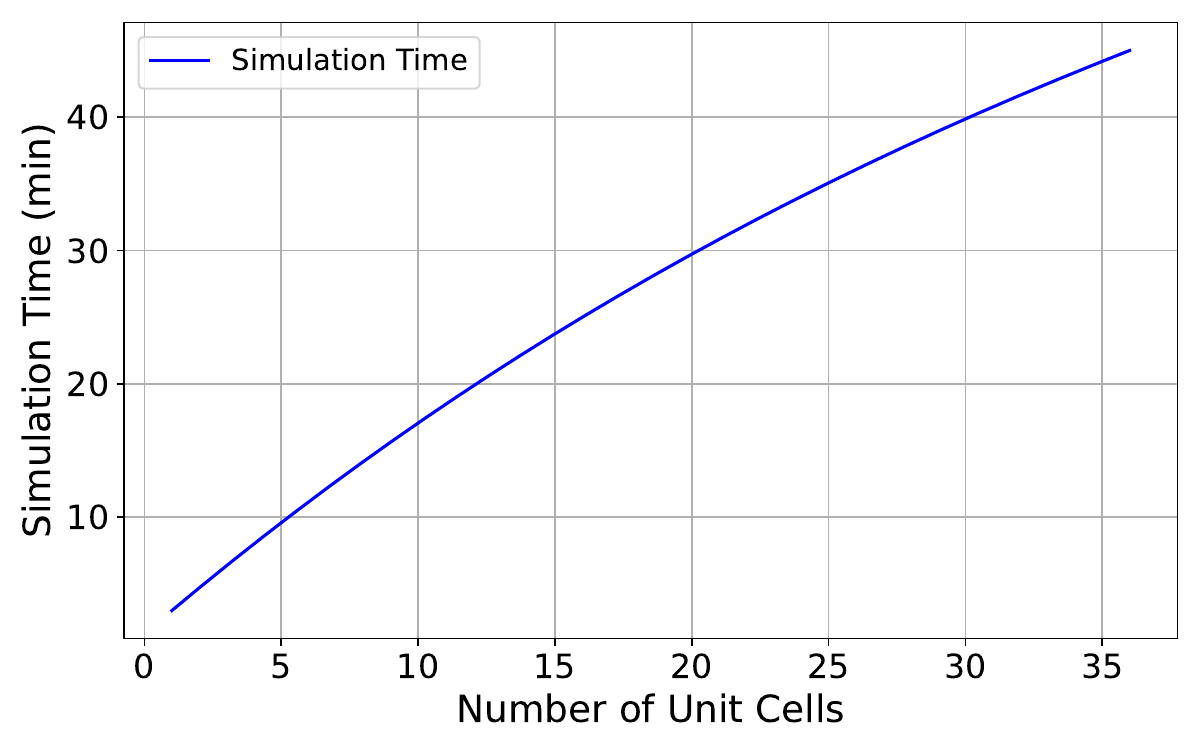}
\caption{A plot showing the computational time required by EM-simulations for the metasurface described in section IV.A with number of unit cells.}
\label{fig:comp_time}
\end{figure}

\subsection{Computational Time}
Our proposed MetaFAP model achieves a dramatic reduction in computational time and memory usage compared to conventional full-wave EM simulations. Specifically, MetaFAP requires only 0.15 ms per inference, while its entire training process completes in less than 30 minutes on an NVIDIA A100 GPU. Our model size is less than 16KB and only has 3857 parameters in total. In contrast, full-wave EM simulation methods typically take on the order of 20–30s per single resonator simulation. Often, they can extend to several hours or even days when applied to large-scale metasurface arrays—especially considering the need for extensive parameter sweeps and iterative optimization \cite{simulation_time}. This stark difference is a consequence of the inherent computational complexity of solving Maxwell’s equations in full-wave solvers, as well as the significant memory overhead associated. Our approach reduces simulation time by up to two orders of magnitude and memory consumption by more than two orders of magnitude. Figure \ref{fig:comp_time} shows that full-wave simulations become significantly slower with increasing unit cells. In contrast, MetaFAP maintains a constant inference time of just 0.15 ms, independent of metasurface size. Clearly, MetaFAP achieves substantial speed and efficiency improvements compared to traditional methods. Hence, it enables rapid and scalable predictions, offering significant computational advantages for real-time RIS applications.

\section{Discussion}

Our experimental analysis demonstrates that MetaFAP significantly surpasses traditional ML and DL methods in accurately predicting metasurface behavior. This enhanced performance primarily stems from two key strengths inherent to meta-learning: a robust initialization that provides an effective starting point for adapting to new frequency ranges, and a rapid fine-tuning capability requiring minimal additional data. These advantages sharply contrast with conventional deep learning models, such as IR-DNN, which struggle with severe overfitting and poor extrapolation beyond their training data.

The practical implications for metasurface design are profound. In next-generation wireless systems—such as 6G or dynamic mmWave and THz networks—RIS must adapt swiftly and accurately to varying spectral conditions, user mobility, or environmental changes. For instance, consider a 6G scenario involving user mobility in urban settings, where devices frequently transition between sub-6 GHz and mmWave frequencies. Traditional EM simulation methods are impractical here due to their computational demands, rendering real-time adaptation unfeasible. However, MetaFAP's capability to provide predictions within 0.15 ms makes real-time adaptive beamforming and interference mitigation practical, significantly enhancing the reliability and quality of service.

Similarly, in intelligent transportation systems (ITS) utilizing RIS-enabled vehicle-to-everything (V2X) communication, rapid frequency adaptation is crucial due to the highly dynamic nature of vehicular environments. MetaFAP's low-latency inference allows for instant RIS adjustments, facilitating continuous and stable communication even as vehicles move through diverse propagation environments or shift frequency bands due to spectrum availability constraints.

Additionally, large-scale RIS deployments in smart city infrastructures—such as RIS-assisted IoT networks—face challenges in scalability and computational efficiency \cite{metasurface_discussion_largescale}. MetaFAP's modest memory requirements (under 16KB model size) and short training duration (less than 30 minutes) directly address these constraints, enabling efficient network-wide RIS configuration updates without heavy computational resources.

Beyond these direct applications, our findings highlight the broader potential of meta-learning in electromagnetic surrogate modeling. Integrating meta-learning with physics-informed constraints could further enhance predictive models' robustness, particularly for complex, multi-resonator coupling scenarios that are common in practical metasurface designs. Future research directions include developing advanced adaptive few-shot methods that dynamically select the most informative data points during real-time operation, further improving prediction accuracy and adaptability in highly variable real-world conditions.

\section{Conclusion}

In this paper, we introduced MetaFAP, a novel meta-learning framework based on the MAML paradigm specifically tailored for frequency-agnostic prediction of metasurface properties. MetaFAP learns a robust initialization across diverse frequency tasks, enabling rapid and accurate adaptation to previously unseen spectral conditions with minimal data. Experimental results confirm that MetaFAP significantly outperforms conventional ML and DL models. Our approach achieved an order-of-magnitude improvement in prediction accuracy with MSE and MAE as low as 0.0079 and 0.0612, respectively. This framework maintains high Pearson correlation coefficients of approximately 80\% across diverse datasets. Furthermore, our approach
offers dramatic reductions in computational overhead, with
inference times as low as 0.15 ms. This starkly contrasts full-wave
electromagnetic simulations that require minutes per unit
cell and scale non-linearly with array size. MetaFAP's remarkable computational efficiency directly facilitates practical applications such as real-time dynamic beamforming in frequency-agile networks and adaptive interference management in complex environments. All in all, MetaFAP establishes a new benchmark for metasurface property prediction, significantly surpassing previous state-of-the-art ML and DL-based methods.

\bibliographystyle{IEEEtran}
\bibliography{main_bib} 

\end{document}